\documentclass[twocolumn,aps,draft]{revtex4}
\usepackage{graphicx}
\usepackage{epsf}

\begin{document}

\title{Alternative Interpretation of Sharply Rising E0 Strengths in 
Transitional Regions}

\author{P. von Brentano$^1$, V. Werner$^1$, R.F. Casten$^{1,2,3}$, 
C. Scholl$^1$, E.A. McCutchan$^2$, R. Kr\"{u}cken$^3$ and J.Jolie$^1$}
\affiliation{$^1$Institut f\"{u}r Kernphysik,
Universit\"{a}t zu K\"{o}ln, K\"{o}ln, GERMANY\\
$^2$Wright Nuclear Structure Laboratory, Yale University, New
Haven Connecticut 06520-8124, USA\\
$^3$Physik Department E12, Technische Universi\"{a}t M\"{u}nchen,
85748 Garching, GERMANY}

\begin{abstract}
It is shown that strong 0$^+_2 \rightarrow$ 0$^+_1$ E0 transitions
provide a clear signature of phase transitional behavior in finite
nuclei. Calculations using the IBA show that these transition strengths
exhibit a dramatic and robust increase in spherical-deformed shape
transition regions, that this rise matches well the existing data,
that the predictions of these E0 transitions remain large in deformed
nuclei, and that these properties are intrinsic to the way that
collecitvity and deformation develop through the phase transitional
region in the model, arising from the specific $d$-boson coherence in
the wave functions, and that they do not necessarily require the
explicit mixing of normal and intruder configurations from different
IBA spaces.
\end{abstract}
\maketitle

Phase transitions are a fundamental feature of many physical systems
and have recently been of considerable interest, including extensive
studies in atomic nuclei
\cite{War02,Abb00,QGP,Ell02,Liu01,LoC96,Iac00,Iac01,Cas98,Cej00,Jol01,Jol02,Ari03,Ari01,12,Zam02,Cej03}
and in other mesoscopic systems, {\it e.g.} \cite{Iac03}. One very active
area of study has been the study of shape changes at low energy in
nuclei
\cite{LoC96,Cas98,Iac00,Iac01,Cej00,Jol01,Jol02,Ari03,Ari01,12,Zam02,Cej03,Iac04}
which have been described using catastrophe \cite{LoC96} and
Landau theory \cite{Jol02,Cej03,Iac04}. Thus far, such studies have
focused on data and 
model comparisons for energies, E2 transition matrix elements, and
quadrupole moments. However, there has been little study of E0 matrix
elements in these shape transitional regions despite the fact that,
since such shape changes are inherently linked to changes in nuclear
shapes and radii, the E0 operator and its transition matrix elements
should provide a fundamental measure of how these phase transitions
proceed \cite{1}.

To study this problem, we exploit the IBA model \cite{2}, which provides an
economic and convenient approach to study both phase transitional
behavior, in which a simple two term Hamiltonian of Ising type
describes transitional regions in terms of variations of a single
control parameter, and E0 transitions, for which the $d$-boson content
is explicitly related to the deformation through the intrinsic state
formalism \cite{2,3,4}.

There have, of course, been some studies of E0 transitions in
transitional nuclei, most notably in the context of the IBA in the
early work of Scholten \textit{et al.} \cite{5}. Their calculations for
Sm isotopes provided anecdotal 
(\textit{i.e.}, parameter-specific) evidence for an increase in E0
strength in deformed nuclei. Large values are also indicated in
analytic expressions for $\rho^2$(E0;0$^+_2 \rightarrow$ 0$^+_1$) 
values in the O(6) and SU(3) limits \cite{1,2,6,7}. Estep
\textit{et al} \cite{8} used calculations from ref. \cite{9} in a
shape coexistence formalism \cite{10} to predict $\rho^2$(E0;0$^+_2
\rightarrow$ 0$^+_1$) values in the Mo isotopes (see below).

However, it is the purpose of this Letter to approach the question of E0
transitions in transitional nuclei in a much more general way,
focusing on generic properties of $\rho^2$(E0) values. We will
use a simple but general Hamiltonian to span the full symmetry
triangle \cite{11} of the IBA and will display complete
contours of these monopole transitions that reveal robust,
parameter-free characteristics of the model. As might be expected
from our earlier comments, the most interesting behavior occurs
precisely in shape transition regions, namely one finds a very
sharp increase in $\rho^2$(E0;0$^+_2 \rightarrow$ 0$^+_1$),
which then \textit{remains} large for well-deformed nuclei.  We
will show that, contrary to common opinion, this characteristic
behavior of 0$^+_2 \rightarrow$ 0$^+_1$ E0 transition strengths
does $not$ require an explicit mixing of coexisting spherical and
deformed intruder configurations. Rather, it arises from a mixing
of components with different $d$-boson content which is a natural
ingredient in the IBA when U(5) symmetry is broken. By analyzing
the calculated $\rho$(E0) matrix elements in terms of
contributions with different n$_d$ values, we will show the key
role of the $d$-boson coherence in the wave functions and that,
while large n$_d$ values are a necessary condition for large
$\rho^2$ values, they are definitely not a sufficient condition.
Finally, while surprisingly little data exists on 0$^+_2
\rightarrow$ 0$^+_1$ E0 transitions, much that does exist happens
to be in shape transitional regions and we will see that the
robust IBA predictions agree with these data.

We start with a simple IBM-1 Hamiltonian \cite{12} that includes
spherical-and deformation-driving terms whose competition
determines the resulting structure,
\begin{equation}
\label{eq:hecqf}
H = a \ \left[(1 - \zeta) n_d - \frac{\zeta}{4N_{B}} Q \cdot Q\right] \ ,
\end{equation}
where Q = s$^\dag\widetilde{d}$ + d$^\dag$s + $\chi$
(d$^\dag\widetilde{d}$)$^{(2)}$ with $\chi \in$
[-$\sqrt{7}$/2, 0].  For $\zeta$ = 0 one obtains the U(5) limit
while $\zeta$ = 1 and $\chi$ = -$\sqrt{7}$/2 gives SU(3), and
$\zeta$ = 1 and $\chi$ = 0 gives O(6). In general,  there is a
spherical-deformed first order phase transition as a function of
$\zeta$ (except for $\chi$ = 0 where it is second order). The
transition is most abrupt for $\chi$ = -$\sqrt{7}$/2 and occurs at
$\zeta$ = 0.5 for large N, and at $\zeta \sim$ 0.54 for
typical boson numbers (N $\sim$ 10). The
E0 transition operator is \cite{1,5}
\begin{equation}
\rho(E0) = \alpha (s^\dag s)^0 + \beta (d^\dag \widetilde{d})^0 =
\alpha N + \beta{^\prime} (d^\dag \widetilde{d})^0 \ .
\end{equation}
The first term vanishes for transitions and the
connection to $d$-boson content is obvious. The results of our
calculations, spanning the entire parameter space for N = 4, 10,
and 16 are shown in Fig. 1A.

The essential result is immediately obvious, namely, that
$\rho^2$(E0;0$^+_2 \rightarrow$ 0$^+_1$) rises dramatically just in
the shape transition region, and remains large in deformed nuclei.
This qualitative result is independent of boson number
(\textit{i.e.}, N, Z), and of $\chi$.  That is, there is no
trajectory from spherical to deformed that avoids this increase.
(Only the detailed trajectory of $\rho^2$ depends on how $\zeta$
and $\chi$ vary.) This is a robust, parameter-free prediction of
the model, inherent to its structure.

These large E0 transitions in the IBA raise a side issue that
would be worth further exploration. The bosons in the IBA
correspond to correlated pairs of nucleons in the valence space.
Yet, microscopically, E0 transitions are forbidden in a single
harmonic oscillator shell \cite{1}. However, realistic shell model
descriptions effectively entail mixing of several oscillator
shells, which is reflected in the use of effective charges in
calculations within restricted spaces. The IBA should incorporate
such effects.

The sharp drop in $\rho^2$ for $\chi \rightarrow$ 0 at lower right
(going toward the O(6) limit) in the plots of Fig. 1A occurs
because of a mixing and crossing of the 0$^+_2$ and 0$^+_3$
states.  This is illustrated for N =10 in Fig. 1B (left).
Comparison with Fig. 1A shows that the 0$^+_2$ and 0$^+_3$ E0
strengths interchange, and large 0$^+_3 \rightarrow$ 0$^+_1$
transitions emerge and persist into the O(6) limit where they are
the allowed transition from the $\sigma$ = (N - 2) 0$^+$ state
\cite{1}. If the 0$^+_3 \rightarrow$ 0$^+_1$ and 0$^+_2
\rightarrow$ 0$^+_1$ values are added, the contour plot remains
nearly constant after the phase transition region [Fig. 1B
(right)]. Other than this case, the only strong ground state E0
transition is 0$^+_2 \rightarrow$ 0$^+_1$ although strong
transitions between pairs of excited 0$^+$ states abound.

It is useful to decompose the E0 strengths in terms of individual
components in the wave functions. This is done for N = 10 and
$\chi$ = -$\sqrt{7}$/2 in Fig. 2 which shows, for three $\zeta$
values (one before the transition, one near the critical point,
and $\zeta$ = 1 for a well deformed rotor), the contributions to
$\rho$(E0;0$^+_2 \rightarrow$ 0$^+_1$) from each n$_d$ value.
These are calculated from $\alpha_2$(n$_d$)$\alpha_1$(n$_d$)n$_d$
where $\alpha_{1,2}$(n$_d$) are the amplitudes in the 0$^+_1$ and
0$^+_2$ states with n$_d$ bosons. In U(5), the 0$^+_1$ and 0$^+_2$
states have n$_d$ = 0 and n$_d$ = 2, respectively and hence, by
orthogonality, $\rho$(E0;0$^+_2 \rightarrow$ 0$^+_1$) = 0. With
increasing U(5) symmetry breaking by $\zeta
\rightarrow 1$, n$_d$ is no longer a good quantum number. In
fact, since $\Sigma<$0$^+_{i>1}|$n$_d|$0$^+_1>^2$ = $<$n$_d^2>$ -
$<$n$_d>^2$, the total E0 strength is related to the spreading
(fluctuations) of n$_d$ in the ground state. As higher n$_d$
components grow \cite{2,12,13} so do their contributions to
$\rho^2$. Such $d$-boson mixing is inherently related to the onset
of quadrupole deformation \cite{2,3,4,14}.

Before the phase transition the $\rho^2$ values are dominated by
coherent n$_d$ = 2, 3 and 4 components. After the phase transition
subtle positive and negative cancellations appear. Higher n$_d$
components are essential to the final sum over $\Sigma$ $\alpha_1$
(n$_d$) $\alpha_2$ (n$_d$) n$_d$. While finite $d$-boson amplitudes
are clearly a necessary condition for both deformation and $\rho$
values, large $\rho$ values are not merely a trivial consequence
of large $<$n$_d>$ values. The many small $\rho$(E0;0$^+_i
\rightarrow$ 0$^+_j$) values prove this. This is illustrated in
the last panel of Fig. 2 which clearly shows the cancellations
that give small $\rho$ values for weak E0 transitions. Rather, it
is the \textit{specific d-boson coherence} in the wave functions
that controls the resultant $\rho$ values.

While the focus here is on 0$^+_i \rightarrow$ 0$^+_1$
transitions, we briefly comment on the behavior for higher spin.
Calculations like those in Fig. 1, but for 2$^+_i$ $\rightarrow$
2$^+_1$ and 4$^+_i$ $\rightarrow$ 4$^+_1$ E0 transitions, show
similar behavior if the $\rho^2$ strengths are summed over all
initial states. However, there is more fragmentation. More than
one initial state has E0 strength to the same yrast state for a
given region of the triangle, and different initial states
dominate the E0 decay in different regions. Empirically,
ref. \cite{1} lists a number of strong E0 transitions to the first
2$^+$ state and there seems to be enhanced fragmentation as well.

The robust predictions of $\rho^2$(E0;0$^+_2 \rightarrow$
0$^+_1$) demand experimental testing. E0 0$^+_2 \rightarrow$
0$^+_1$ transitions are known \cite{1} in both the A = 100 and 150
transition regions. Figure 3 compares these data with schematic
IBA calculations, using $\chi$ = -$\sqrt{7}$/2 and N = 10. The
data are plotted at $\zeta$ values where the calculations
reproduce the experimental R$_{4/2}$ values. Despite the
restriction to a fixed $\chi$, a constant boson number N,
and that $\zeta$ was chosen simply by fitting two yrast energies,
these calculations clearly reproduce the sharp rise in $\rho^2$(E0)
values.

These results raise an important question relating to phase
transitional behavior. Microscopically, the Federman-Pittel
mechanism \cite{15}, which invokes strong p-n interactions
\cite{16}, leads to single particle energy shifts (via the
monopole component \cite{17}) and to the descent of a coexisting
deformed configuration in otherwise spherical nuclei. An
equilibrium deformation ensues when this configuration becomes the
ground state. In the IBA, this coexistence can be explicitly
included by the Duval-Barrett formalism \cite{10} in which a pair
of nucleons (protons in this case) is excited across a shell or
subshell gap to form a space with N$_{\pi_{def}}$ =
N$_{\pi_{sph}}$ + 2 (counting the extra pairs of holes and
particles as additional bosons), thus, H = H$_{N_{B}}$ +
H$_{N_{B}+2}$ + H$_{mix}$. Typical Duval-Barrett calculations
involve many parameters -- two or more for each term in H. The
calculations of ref. \cite{9} used 13 parameters but reproduce the
experimental $\rho^2$(E0;0$^+_2 \rightarrow$ 0$^+_1$) values
(see Fig. 3) in Mo rather well.

The interesting point, however, is that, while the large $\rho^2$
values in these calculations have been ascribed \cite{8} to the
mixing, that is, to a non-vanishing H$_{mix}$ (since, without
H$_{mix}$, E0 transitions between states of H$_{N_{\pi=1}}$ and
H$_{N_{\pi=3}}$ are forbidden), it is evident from Figs. 1 and 3
that large values of $\rho^2$(E0;0$^+_2 \rightarrow$ 0$^+_1$)
$also$ occur in the IBA $without$ the need to introduce such
mixing.

How can these seemingly conflicting results be reconciled? Figure
5 of ref. \cite{9} shows the probabilities of $N_{\pi=1}$ and
$N_{\pi=3}$ components in the ground state wave functions (and, by
orthogonality, the approximate admixtures for the 0$^+_2$ states).
There is, in fact, little mixing ($<$10$\%$) for N = 54 (spherical
Mo nuclei) and even less ($<$5$\%$) for N = 60, 62 (the first and
second deformed Mo isotopes). $Only$ for N = 56, 58 is there
substantial mixing. Thus, these Duval-Barrett calculations
effectively go over into the simple (single space) IBA results
before and after the transition region. It is therefore $not$ the
large $\rho^2$ value for N = 60 that requires mixing. It is rather
the $moderate$ $\rho^2$ values for the {\it pre-deformed} transitional
Mo isotopes with N = 56, 58. This interpretation is validated by
other observables. In $^{96}$Mo$_{54}$, $^{98}$Mo$_{56}$, the
experimental values of the ratios B(E2;0$^+_2 \rightarrow$
2$^+_1$)/B(E2;2$^+_1 \rightarrow$ 0$^+_1$) and B(E2;2$^+_2
\rightarrow$ 2$^+_1$)/B(E2;2$^+_1 \rightarrow$ 0$^+_1$) $exceed$
any predictions of standard models, including the vibrator and
rotor. The reason is that the 0$^+_1$ state primarily consists of
N bosons while the 2$^+_1$, 0$^+_2$ and 2$^+_2$ states belong
primarily to the N + 2 space \cite{9}. Hence, the denominators
are hindered. It requires the Duval-Barrett formalism with
parameterized H$_{mix}$ to account for these data.

The key point here is that large $\rho^2$(E0;0$^+_2 \rightarrow$
0$^+_1$) values in transitional nuclei can arise in two ways,
$either$ from mixing of coexisting spherical and intruder
configurations (as shown in ref. \cite{7}) originating in
different spaces (see ref. \cite{18}), $or$, alternately, from the
simpler IBA-1 itself. Thus, contrary to many statements in the
literature, strong spherical-intruder state mixing is not
$required$ for large $\rho^2$(E0) values, $nor$ are large
experimental $\rho^2$(E0) values in transitional nuclei
$necessarily$ a signature of such mixing effects. One must analyze
each region to determine whether to explicitly introduce shape
mixing or whether the simple, few parameter, IBA alone suffices.
In the Mo region ref. \cite{9} shows that mixing of
shape coexisting states is essential for the pre-deformed nuclei.
However, in the first deformed nuclei in both the mass 100
($^{98}$Sr, $^{100}$Zr, $^{102}$Mo) and 150 ($^{152}$Sm,
$^{154}$Gd) regions some of the largest known $\rho^2$(E0) values
are easily accounted for $without$ such mixing by the IBA-1.

We commented above that E0 transitions vanish in a single oscillator
shell. It is therefore of interest to study how a valence space model
such as the IBA, in which the bosons are considered to be formed from
nucleons in the first open shell beyond an inert doubly magic core,
can produce large E0 strengths. Of course, E0 transitions can arise by
coupling to the giant monopole resonance, but this would seem to be
outside the IBA space. Rather, the E0 transitions in the IBA may
reflect the fact that realistic major shells in the independent
particle model include an intruder orbit from the next higher shell,
and that additional intruder orbits, from both lower and higher
shells, appear in the Nilsson scheme with increasing deformation, that
is, as the phase transition proceeds. Of course, as a phenomenological
model, one cannot relate the IBA directly to such a picture without
detailed microscopic analysis, but it may be that the importance of
intruder orbits is reflected in the effective parameter, $\beta'$, in
the E0 operator, which, in the calculations presented in Fig. 3, was
fixed at $6\cdot 10^{-3}/eR_0^2$. Remarkably, the use of a simple
one-body operator with constant coefficients is sufficient for
reproducing the trends of the data in transition regions. Given the
empirical success shown here for the simple IBA interpretation of E0
transitions, microscopic studies are strongly encouraged.

Lastly, one upshot of this study concerns well-deformed nuclei.
The only $\rho^2$(E0;0$^+_2 \rightarrow$ 0$^+_1$) values known
in the deformed rare earth nuclei are very small values ($\rho^2
\sim 2\cdot 10^{-3}$) in $^{166}$Er and $^{172}$Yb, in contrast to the IBA
predictions. However, the empirical 0$^+_2$ states may not
correspond to the 0$^+_2$ states of the IBA, but could have
two-quasi-particle character. Interestingly, in the neighboring
nucleus $^{170}$Yb, there is a rather strong E0 transition
($\rho^2 = 27(5)\cdot 10^{-3}$) from the 0$^+_3$ state to the ground state. It
is also interesting that there are a number of large
$\rho^2$(E0;2$^+_i \rightarrow$ 2$^+_1$) values known for deformed
nuclei \cite{1}. Moreover, in recent IBA calculations \cite{19},
anomalous kinks in the parameter systematics are avoided if the
empirical 0$^+_3$ state is associated with the 0$^+_2$ IBA state
near A = 170. Clearly, it is important to measure 0$^+_i
\rightarrow$ 0$^+_1$ E0 transitions in a number of deformed nuclei
to see if the total E0 strength predicted in the IBA is recovered.

To summarize, experimentally, $\rho^2$(E0;0$^+_2 \rightarrow$
0$^+_1$) values rise dramatically in shape/phase spherical-deformed
transition regions. We have presented here an alternative view in
which this rise, and large E0 transitions in deformed nuclei, arise
not from mixing of coexisting spherical and deformed configurations,
although such a mechanism may contribute as well in specific instances
({\it e.g.}, $^{98,100}$Mo), but rather from $\beta$-deformation and
its variation in the transition region. This result is directly
connected to the physics of phase transitional regions since
calculations within a single space reproduce the characteristic
increase in E0 transition strengths. That is, using the IBA-1 model,
we showed that contrary to common opinion, the rise in $\rho^2$(E0)
values is predicted even by this simple, single space model, that it
agrees with the data, is parameter-free and intrinsic to the model,
does not require the mixing of different IBA spaces, and develops due
to the specific $d$-boson coherence in the wave functions. In the IBA-1
the E0 strengths are directly related to the fluctuations (spreading)
in $n_d$ values, and therefore to the $\beta$-deformation. Finally,
we have proposed a direct test of these ideas through the measurement
of E0 transitions to the ground state in well deformed nuclei.

We are grateful to T. Otsuka and F. Iachello for useful
discussions, especially of the n$_d$ content of the wave
functions, and we thank B. Barrett for discussions on the
Duval-Barrett formalism. Work supported by USDOE grant number
DE-FG02-91ER-40609, by DFG contract number Br 799/12-1, and by BMBF
grant number 06K167.

\bigskip

\begin{figure}
\epsfxsize 8.6cm
  \epsfbox{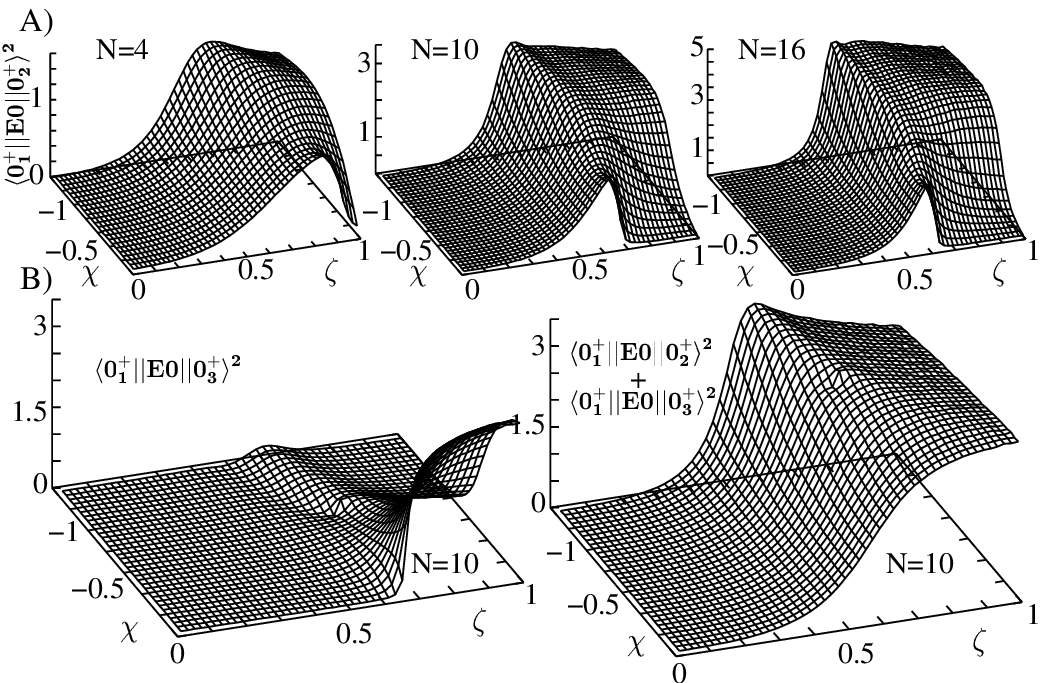}
\caption{A) Contour plots of $\rho^2$(E0;0$^+_2
\rightarrow$ 0$^+_1$) throughout the IBA parameter space for N
= 4, 10, 16. The range of $\chi$ values implicit in the U(5) limit
is explicitly shown along the left axis; B) Contour plots for
N = 10, similar to the top panel, but for $\rho^2$(E0;0$^+_3
\rightarrow$ 0$^+_1$) on the left and for the sum $\rho^2$(E0;0$^+_2
\rightarrow$ 0$^+_1$) + $\rho^2$(E0;0$^+_3\rightarrow$ 0$^+_1$) on
the right.}
\end{figure}%

\bigskip

\begin{figure}
\epsfxsize 8cm
  \epsfbox{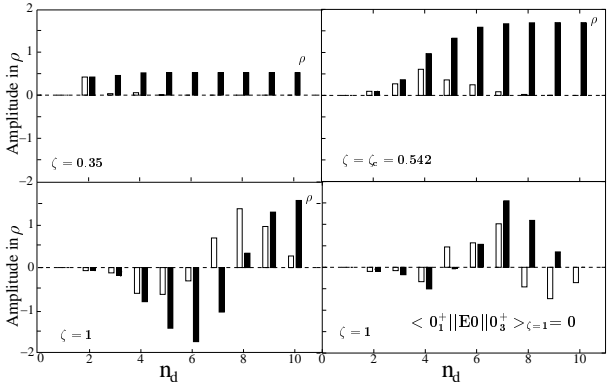}
\caption{Decomposition of the $\rho$(E0) amplitudes as
a function of n$_d$ for the 0$^+_2 \rightarrow$ 0$^+_1$ transition
for $\zeta$ = 0.35, 0.54 (near the critical point), and $\zeta$ =
1 and for 0$^+_3 \rightarrow$ 0$^+_1$ for $\zeta$ = 1 at lower
right, for N = 10, $\chi$ = -$\sqrt{7}$/2. Open bars are
amplitudes for a given n$_d$ and solid bars are the running sum
from n$_d$ = 0 up to the given n$_d$ value.}
\end{figure}%

\bigskip

\begin{figure}
\epsfxsize 8cm
  \epsfbox{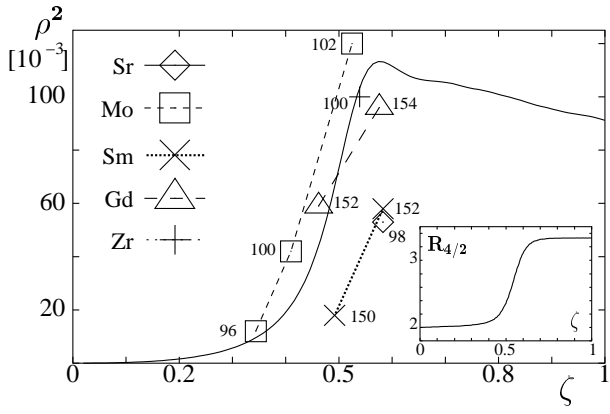}
\caption{Empirical $\rho^2$(E0;0$^+_2 \rightarrow$
0$^+_1$) values (from ref. \cite{1}) for nuclei in the A
= 100 and 150 transition regions and schematic IBA-1
calculations. (As such schematic calculations cannot give
R$_{4/2}<$2, nuclei such as $^{98}$Mo are not considered.)
The solid curve is the IBA prediction for N = 10 with $\chi$ =
-$\sqrt{7}$/2 and $\beta^\prime$ (eq. 2) = $6 \cdot
10^{-3}/eR_0^2$. The inset shows how R$_{4/2}$ itself behaves with
$\zeta$: note the similarity to the $\rho^2$(E0) trajectory. Data
points are labelled with mass number A.}
\end{figure}%

\end{document}